\documentclass[usenatbib]{mnras}
\usepackage[utf8]{inputenc}
\usepackage{graphicx}
\usepackage{multirow}
\usepackage{amsmath}
\usepackage{amssymb}
\usepackage{hhline}
\newcommand{\be}{\begin{equation}}
\newcommand{\ee}{\end{equation}}

\newcommand{\dq}{\delta_Q}
\newcommand{\dkq}{\tilde{\delta}_Q}
\newcommand{\K}{\mathbf{k}}
\newcommand{\X}{\mathbf{x}}
\newcommand{\R}{\mathbf{r}}
\newcommand{\im}{\mathrm{i}}
\newcommand{\cg}{\left<\left|\dkq(\K)\right|^2\right>}
\newcommand{\caL}{\left<\tilde{T}(\K,L_1)\tilde{T}^*(\K,L_2)\right>}
\newcommand{\ca}{\left<\tilde{T}(\K)\tilde{T}^*(\K)\right>}
\newcommand{\cx}{\left<\tilde{T}(\K)\tilde{\delta}_Q^*(\K)\right>}
\newcommand{\cxL}{\left<\tilde{T}(\K,L)\tilde{\delta}_Q^*(\K)\right>}
\newcommand{\nq}{\overline{n}_Q}
\newcommand{\exki}{e^{\im\K\cdot(\X_i-\X_j)}}
\newcommand{\exkp}{e^{\im\K\cdot(\X-\X')}}
\newcommand{\clt}{C_{LT}}
\newcommand{\qlf}{\left.\frac{dn}{dL}\right|_Q}
\newcommand{\LQ}{\overline{L}_Q}

\newcommand{\tco}{\overline{T}_{\rm{CO}}}
\newcommand{\subCO}{_{\rm{CO}}}

\begin{document}
\title{Observing AGN feedback with CO intensity mapping}
\author[Breysse \& Alexandroff]{Patrick C. Breysse$^{1}$\thanks{pcbreysse@cita.utoronto.ca} and Rachael M. Alexandroff$^{1,2}$\\
$^{1}$Canadian Institute for Theoretical Astrophysics, University of Toronto, 60 St. George Street, Toronto, ON, M5S 3H8, Canada \\
$^{2}$Dunlap Institute for Astronomy and Astrophysics,University of Toronto, 50 St. George Street, Toronto, ON, M5S 3H4, Canada }

\maketitle

\begin{abstract}
Current models of galaxy formation require star formation in high-mass galaxies to be limited by poorly understood mechanisms of quasar feedback.  Feedback processes can be studied by examining the molecular gas content of AGN hosts through the CO rotational ladder, but the complexity of these observations means that current data are limited to only extremely CO-bright objects. Upcoming CO intensity mapping experiments offer an opportunity for a less biased probe of quasar feedback.  By correlating intensity maps with spectroscopic AGN surveys, we can obtain a measurement of the mean CO luminosity of a large population of quasars simultaneously.  We show that experiments like COMAP, CCAT-prime, and CONCERTO have enough sensitivity to detect this cross-correlation if existing AGN observations are representative of the whole population, and to place interesting upper limits if they are not.  Future surveys will be able to increase the precision of these measurements by orders of magnitude, allowing detailed studies of quasar properties across a wide range of cosmic history.
\end{abstract}


\section{Introduction}
In order for analytic simulations of galaxy formation to match the observed galactic number density over cosmic time, it is necessary to invoke a process to limit the maximal stellar mass of galaxies in the universe \citep{Springel2005, Bower2006, Croton2006, Silk2012}.  At the most massive end of the galaxy mass function, this process is believed to be feedback from the active galactic nuclei (AGN) \citep{Hopkins2006,Fabian2012} which should shut off and prevent further star formation in the host galaxy \citep{Sanders1988, Hopkins2006}.  One tracer of the star formation in galaxies is their molecular gas content. Molecular gas provides the material for star formation and is directly correlated with a galaxy's star formation rate and its ability to turn cold gas in to stars.  Thus, it is crucial to understand how AGN-driven outflows are affecting the molecular gas in their host galaxies.   

 Tracing molecular gas is difficult, however, because the bulk of the molecular gas in the interstellar medium (ISM) is in the form of molecular hydrogen (H$_2$), which has no emission or absorption features at the relatively cool temperatures expected in the ISM (though quasar-driven winds might enhance the emission of warm H$_2$, \citealt{Hill2014}). Instead, carbon monoxide (CO) is a common and useful tracer of the molecular gas content, with several rotational transitions at the correct temperature.

Detections of individual AGN in CO remain relatively scarce and an unbiased study of the molecular gas content of AGN hosts is even more elusive. Often individual objects are targeted because of previous submillimeter detections, their intrinsic bolometric luminosity \citep[e.g.][]{Hill2018} or because of the presence of molecular outflows in other lines \citep[e.g.][]{Cicone2014}. Even with the most sensitive instruments available today, sampling a large area remains difficult. In addition, observations of the molecular gas content of powerful quasars using sub-millimeter continuum and molecular line emission provide some conflicting results. Some objects appear gas-rich and strongly star-forming \citep{Beelen2004,Coppin2008,Huseman2017,Banerji2017} and others depleted of gas \citep{Cicone2014,Kakkad2016,Fiore2017,Carniani2017}. 

Upcoming line intensity mapping observations offer a possible solution to this lack of statistics.  Rather than aiming to image individual known galaxies, intensity maps measure the total emission in a particular spectral line with relatively coarse angular resolution \citep[see][for a review]{Kovetz2017}.  By observing at many closely-spaced frequency bands, intensity mapping surveys can probe structure in three dimensions with sub-percent redshift accuracy.  An intensity map thus probes the combined emission of all sources within a given volume, including faint objects which would be difficult or impossible to target individually.  While the technique cannot tell us the line intensity of any individual object, it is sensitive to the statistical properties of all emitting galaxies, and can be used to make many useful inferences about the nature of high-redshift galaxies \citep[see, e.g.,][]{Breysse2016a,Comaschi2016,Serra2016,Lagache2018}.

Though intensity mapping is a relatively young field, it has attracted significant experimental investment.  Intensity mapping of the 21 cm spin-flip transition in neutral hydrogen has long been seen as a useful technique for studying cosmology \citep[][and references therein]{Pritchard2012}, with several surveys completed or in progress across a wide range of redshifts \citep{Tingay2013,vanHaarlem2013,Bandura2014,Ali2015,Xu2015,Newburgh2016,DeBoer2017}, and a pair of detections in cross-correlation \citep{Masui2013,Anderson2018}.  In addition to this, there has been a surge of recent interest in intensity mapping of other lines, including fine-structure lines of \ion{C}{II} \citep{Gong2012,Yue2015} and other hydrogen transitions such as H$\alpha$ and Ly$\alpha$ \citep{Dore2014,Pullen2014,Silva2018,Cooray2016} among others.  Most interesting for our purposes, however are experiments targeting rotational CO transitions.

The CO intensity mapping signal was first modeled in \citet{Righi2008}, who were primarily interested in it as a possible foreground for cosmic microwave background observations.  It was quickly realized that CO fluctuation maps were interesting in their own right as tracers of molecular gas and star formation in high-redshift galaxies \citep{Lidz2011,Pullen2013,Breysse2014}.  A tentative detection of small-scale CO(1-0) intensity fluctuations at $z\sim3$ was reported by the CO Power Spectrum Survey (COPSS) using inteferometric observations \citep{Keating2015,Keating2016}.  Currently, the CO Mapping Array Pathfinder (COMAP) experiment is underway which will have sufficient sensitivity for a high-significance detection over a similar redshift range and a much larger volume. “Phase II” of COMAP will target the same redshift at higher sensitivity \citep{Li2016} and a planned  “Phase III” will add a second frequency channel to target z=6-8 at the end of the Epoch of Reionization (EoR).

Surveys which do not have CO as their primary target can also be useful for studying molecular gas in AGN.  Specifically, several higher-$J$ CO transitions between $z\sim0-2$ fall into the frequency ranges of experiments targeting the \ion{C}{II} line at the EoR.  These lines are typically treated as foreground contaminants for reionization science, and substantial effort has gone into studying how they can be subtracted from \ion{C}{II} maps \citep{Breysse2015,Cheng2016,Lidz2016,Sun2018}.  However, these so-called ``foregrounds" carry important astrophysical information in their own right.  We will consider here three \ion{C}{II} experiments which are planned to take data over the next few years: the Tomographic Ionized-Carbon Intensity Mapping Experiment (TIME, \citealt{Crites2014}), the Cerro Chajnantor Atacama Telescope-prime (CCATp, \citealt{Stacey2018}), and the Carbon \ion{C}{II} line in post-reionization epoch (CONCERTO, \citealt{Lagache2018}).  We will also consider a hypothetical futuristic \ion{C}{II} experiment which has been termed \ion{C}{II} Stage II (abbreviated StageII hereafter) defined in \citet{Silva2015}.  An additional \ion{C}{II} experiment, the Experiment for Cryogenic Large-aperture Intensity Mapping (EXCLAIM) has recently been announced \citep{Padmanabhan2018}, but they target much lower redshift than the other experiments we consider here ($z\sim3$ for \ion{C}{II} rather than $z\sim6-8$), and as such the CO lines they are sensitive to will be either higher-$J$ or lower redshift.  We therefore leave detailed forecats of EXCLAIM sensitivities to future work.

As stated previously, intensity maps are sensitive to the total emission from all emitters in a surveyed volume.  In order to specifically study AGN, we need a way to isolate the contribution from AGN to the total emission.  This can be done using cross-correlation.  \citet{Wolz2017} (W17 hereafter) showed that, on small scales, the cross-power spectrum\footnote{This same statement can be expressed mathematically equivalently in terms of a stacking analysis, or in terms of the real-space two point function.  However, as the intensity mapping literature predominantly prefers to work with power spectra, we choose to follow W17 and express our forecasts in Fourier space.} between a line intensity map and a galaxy survey depends on the mean line luminosity of only those sources contained in the galaxy catalog.  This means that by cross-correlating a CO intensity map and a map of quasar positions we can directly target the molecular gas contained within the mapped AGN.  In doing so, we will be able to simultaneously observe dozens or hundreds of AGNs without many of the selection biases that plague targeted molecular gas observations.  It is this powerful technique that will allow for a systematic study of the molecular gas content of AGN hosts over large portions of cosmic time.

This paper is organized as follows: Section \ref{sec:method} describes our cross-correlation technique and how it can be used to forecast constraints on molecular gas in AGN hosts, then proceeds to give the parameters of the fiducial surveys we consider. Section \ref{sec:results} gives the results of our forecasts, and we describe the science implications of these forecasts in Section \ref{sec:science}.  Further discussion is found in Section \ref{sec:discussion}, and we conclude in Section \ref{sec:conclusion}.

\section{Method}
\label{sec:method}
Cosmological density fields, whether they be maps of galaxy density or line intensity, are often described in terms of their power spectra.  For a map of the number density $n_Q(\X)$ of AGN, we typically consider the power spectrum of the dimensionless density contrast $\delta_Q\equiv(n_Q(\X)-\overline{n}_Q)/\overline{n}_Q$.  As quasars are a biased, discrete tracer of the underlying dark matter density field, we can write this power spectrum as
\be
P_Q(k)=b^2_QP_m(k)+\frac{1}{\overline{n}_Q},
\label{galauto}
\ee
where $b_Q$ is the quasar bias and $1/\overline{n}_Q$ is the shot noise caused by Poisson randomness in the galaxy positions.  For the convenience of the reader, detailed derivations of this and the other power spectra discussed in this section are provided in Appendix \ref{app:Pk_derivation}.

Qualitatively, the first term in Eq. (\ref{galauto}) comes from the correlation between pairs of galaxies, while the shot-noise term comes from the correlation of each galaxy with itself.  In the intensity mapping regime, we weight each galaxy by its intensity in a chosen line, and take the power spectrum of the resulting intensity field, which we define here in brightness temperature units $T(\X)$.  The auto-spectrum of a map of CO is given by
\be
P_{\rm{CO}}(k)=\overline{T}_{\rm{CO}}^2b_{\rm{CO}}^2P_m(k)+P_{\rm{CO}}^{\rm{shot}},
\label{COauto}
\ee
(see, for example \cite{Lidz2011,Breysse2014}).  The CO emitting galaxies have their own bias $b_{\rm{CO}}$, and the large-scale clustering term is weighted by the sky-averaged mean intensity of the line $\tco$.  There is again a scale-independent shot noise component, as the CO emission is still coming from a population of discrete emitters.

If we assume a luminosity function $dn/dL$ for the CO emitters, we can write the mean intensity as
\be
\tco=\clt\int L\frac{dn}{dL}dL,
\label{TCO}
\ee
For compactness, we have defined the conversion factor between CO luminosity density and observed brightness temperature at redshift $z$ as
\be
\clt=\frac{c^3(1+z)^2}{8\pi k_B\nu\subCO H(z)},
\label{CLT}
\ee
where $c$ is the speed of light, $k_B$ is Boltzmann's constant, $H(z)$ is the Hubble parameter, and $\nu\subCO$ is the rest frequency of the target CO transition.  The CO bias is conventionally defined as
\be
b\subCO=\frac{\int Lb(L) dn/dL dL}{\int Ldn/dL dL},
\label{bCO}
\ee
where $b(L)$ is the bias of a galaxy with CO luminosity $L$.  The shot noise again comes from the correlation of a given CO emitter with itself, and so is proportional to the mean-square CO luminosity. In full,
\be
P_{\rm{CO}}^{\rm{shot}}=\clt^2\int L^2\frac{dn}{dL}dL.
\label{COshot}
\ee
Equations (\ref{TCO}), (\ref{bCO}), and (\ref{COshot}) can be modified to be written in terms of the halo mass function $dn/dM$ by assuming a mass-luminosity relation L(M).

Finally, we can write down the cross-spectrum we would obtain by correlating these two maps:
\be
P_\times(k)=\tco b\subCO b_Q P_m(k)+\clt\LQ,
\label{Px}
\ee
where
\be
\LQ=\frac{1}{\overline{n}_Q}\int L\left.\frac{dn}{dL}\right|_QdL
\ee
is the mean CO luminosity of the quasars, which are assumed to have a CO luminosity function $\left.dn/dL\right|_Q$.  The origin of the first term is clear, each pair of galaxies contributes one factor of $\tco b\subCO$ and one factor of $b_Q$.  The cross-shot power comes from galaxies which simultaneously emit CO and host AGN, and is thus proportional to the mean CO luminosity $\LQ$ of these and only these galaxies.  In other words, a measurement of the shot power in a CO-AGN cross-spectrum provides a measure of the mean CO luminosity of the targeted quasars.  As with the prior auto-spectra, a derivation of Eq. (\ref{Px}) can be found in Appendix \ref{app:Pk_derivation} (also see Eqs. (13) and (14) of W17).

In all of our power spectra thus far we have assumed a linear, scale-independent bias for all tracers.  This approximation is valid on large scales, but deviations from this simple biasing scheme will likely be important on the scales where the power spectrum transitions from clustering to shot-noise dominated (see discussions in W17, \citealt{Chung2019}).  \citet{Wolz2018} demonstrate that halo models can be developed which can fully account for the nonlinear power in simulated intensity maps.  For simplicity here, however, we will neglect these nonlinear contributions, as errors due to this assumption are likely small compared to the order-of-magnitude uncertainties on the modeling of the CO signal.  To be conservative, we will limit our forecasts here to upper limits on $\LQ$, with the understanding that a full measurement would require subtracting a nonlinear component from the cross-spectrum.

We can use the Fisher matrix formalism to estimate how well a given instrument can measure $\LQ$.  Any given survey may not have sufficient resolution to access fully shot-noise dominated scales, so we consider possible degeneracies between $\LQ$ and the amplitude of the large-scale clustering term.  We can write down the covariance matrix of our two density fields as
\be
C_{ij}(k)=\begin{pmatrix} P\subCO(k)+P_N/W^2(k) & P_\times(k) \\ P_\times(k) & P_Q(k) \end{pmatrix},
\ee
where $P_N$ is the noise power spectrum of the CO survey assuming a noise $\sigma_N$ in each voxel.  The only noise in the AGN power spectrum is the shot noise, which we have included in our definition of $P_Q$.  

The quantity $W(k)$ defines the sensitivity cutoff at the resolution limit of the telescope, given by
\be
W(k)=e^{-k^2\sigma_\perp^2}\int_0^1e^{-k^2\left(\sigma_\parallel^2-\sigma_\perp^2\right)\mu^2}d\mu,
\label{Wk}
\ee
where
\be
\sigma_\perp=r(z)\sigma_{\rm{beam}}
\label{sigPerp}
\ee
is the cutoff in the plane of the sky for a beam with width $\sigma_{\rm{beam}}$\footnote{$\sigma_{\rm{beam}}=\theta_{\rm{FWHM}}/\sqrt{8\ln 2}$ for beam full width at half maximum $\theta_{\rm{FWHM}}$.} and a survey at comoving distance $r(z)$.  The cutoff in the line-of-sight is set by the frequency resolution $\delta\nu$, and is given by
\be
\sigma_\parallel=\frac{c}{H(z)}\frac{(1+z)\delta\nu}{\nu_{\rm{obs}}}
\label{sigPar}
\ee
for observed frequency $\nu_{\rm{obs}}$.  Eqs. (\ref{Wk}-\ref{sigPar}) are derived in Appendix C of \citet{Li2016}.

For free parameters $x_i=\left(\tco b\subCO,b_Q,\LQ\right)$, the Fisher matrix is then
\be
F_{ij}=\frac{1}{2}\sum_kN_m(k)\rm{Tr}\left[\frac{\partial C}{\partial x_i}C^{-1}\frac{\partial C}{\partial x_j}C^{-1}\right],
\label{fullFisher}
\ee
\citep{Fonseca2018,Tegmark1997}, where the sum is over bins in $k$ space with width $\Delta k$, each of which contains
\be
N_m=\frac{k^2\Delta k V_{\rm{surv}}}{4\pi^2}
\label{Nmodes}
\ee
modes for a survey covering volume $V_{\rm{surv}}$.  Including both the auto- and cross-spectra in the Fisher matrix allows us to use information in the auto-spectra of the two tracers to break degeneracies in the cross-spectra.

\subsection{Fiducial Surveys}
Here we describe the available AGN maps and the parameters of our planned intensity mapping surveys.

\subsubsection{Quasar Reference Samples}
In order to harness the predictive power of a CO line intensity mapping survey, we require a large sample of quasars for cross-correlation.  While large photometric surveys provide at least an order of magnitude more objects than spectroscopic surveys, the increased redshift uncertainty makes them unsuited for our task \citep[see][]{Chung2019}. Modeling our result with photometric errors is beyond the scope of this paper and will be treated in future work. Instead we focus on current and upcoming deep spectroscopic surveys with sufficient numbers of quasars.

In order to obtain a sufficient number density of current spectroscopically-confirmed AGN, we rely on data from deep field spectroscopic surveys. In particular, the Cosmic Evolution Survey \citep[COSMOS;][]{Scoville2007} field has a wealth of spectroscopic follow-up of X-ray sources, the majority of which are AGN. The entirety of the COSMOS field has been observed by XMM-Newton for a total of 1.55 Ms \citep{Hasinger2007,Cappelluti2007,Cappelluti2009} with the final catalog containing 1887 (point-like) sources detected in either the soft (0.5-2.0 keV), hard (2-10 keV), or ultrahard (5-10 keV) bands. Various surveys \citep{Trump2009,Lilly2009} have conducted spectroscopic follow-up of AGN candidates in this field using IMACS on Magellan \citep{Trump2009} or VIMOS on the VLT \citep[zCOSMOS bright and faint;][]{Lilly2009} which together have covered about half of all AGN-candidates in COSMOS. A catalog of the spectroscopic counterparts to XMM-COSMOS sources is presented in \citet{Brusa2010} which includes a total of 890 unique spectroscopic objects. To construct our sample we include only those objects from \citet{Brusa2010} which are part of the XMM-COSMOS flux limited sample, have only one reliable counterpart with a secure redshift, and are classified by their optical spectroscopy and x-ray detections as being an AGN, not a star-forming galaxy leaving us with a total sample of 730 AGN from XMM-COSMOS.  


Looking towards the future, large, deep, all-sky spectroscopic surveys will allow for a cross-correlation with much larger samples of AGN. For example, the Dark Energy Spectroscopic Instrument (DESI) survey \citep{DESI2016} will operate from the Mayall 4-meter telescope on Kitt Peak over the course of 5 years and observe 14,000 deg$^2$ of the sky. As part of its mandate to study baryon acoustic oscillations (BAO) and redshift-space distortions, DESI will target bright emission line galaxies out to z=1.7, and quasars at higher redshifts ($2.1 < z < 3.5$) with a final sample including approximately 17.1 and 1.7 million objects in each sample respectively. For our purposes, the expected number density of spectroscopic quasars from DESI has been taken from the DESI Science Motivations and Requirements white paper\footnote{https://www.desi.lbl.gov/final-design-report/} Tables 2.3 \& 2.7, depending on the redshift range considered. To derive the total number of quasars expected for a specific CO transition, we interpolate between the numbers provided in the DESI Science Motivations and Requirements document. While the number density of quasars is obviously not going to be as dense as a deep field, DESI will cover a large portion of the sky and therefore large-area \ion{C}{II} experiments will be able to conduct cross-correlations with a very large number of quasars, which will improve their sensitivity.

\subsubsection{CO Surveys}
The surveys we consider here can be broken into two categories.  First we have Phase I and II of the COMAP survey, which target the CO(1-0) transition at $z\sim3$.  Both phases will have some sensitivity to CO(2-1) at $z\sim7$, but as there are few spectroscopic galaxies at such high redshifts we limit ourselves to the $z\sim3$ observations here.  Second, we have the CCATp, CONCERTO, TIME, and \ion{C}{II}-Stage II surveys, which primarily target the \ion{C}{II} transition during reionization.  For these experiments, which we will group together under the general name ``\ion{C}{II} surveys", we are interested here in higher-order CO rotation transitions which are foregrounds to the target \ion{C}{II} line.  As the CO(1-0) and \ion{C}{II} experiments observe in very different frequency ranges, there are different conventions in the literature for computing $P_N$.

For COMAP-type surveys, we typically write
\be
P_N=\sigma_{\rm{vox}}^2V_{\rm{vox}},
\ee
where
\be
\sigma_{\rm{vox}}=\frac{T_{\rm{sys}}}{\sqrt{N_{\rm{det}}\delta\nu t_{\rm{pix}}}},
\ee
is the noise in a single survey voxel, $T_{\rm{sys}}$ is the system temperature, $N_{\rm{det}}$ is the number of detector feeds, and $t_{\rm{pix}}$ is the time spent observing each pixel \citep{Li2016}.  A ``pixel" here is defined to cover a solid angle $\Omega_{\rm{pix}}=\sigma^2_{\rm{beam}}$, so for a survey area $\Omega_{\rm{surv}}$, $t_{\rm{pix}}= t_{\rm{obs}}(\Omega_{\rm{pix}}/\Omega_{\rm{surv}})$.  The volume $V_{\rm{vox}}$ of a single voxel can then be set by separating each 2D pixel into voxels with line-of-sight width set by $\delta\nu$.  

For COMAP Phase I, we consider correlations with the 1.7 deg$^2$ XMM-COSMOS AGN survey, while we consider correlating Phase II with the much larger DESI survey. Table \ref{COMAP_par} gives the parameters we assume here for COMAP Phases I and II.  These are slightly modified from those given in \citet{Ihle2019}.  For Phase I, we reduce the observing time by $1.7/2.25$, the ratio between the full COMAP survey area and that of the COSMOS field.  For Phase II, the full survey plans to target four separate fields, and since we are correlating with DESI we assume that there will be quasar data available in each of them.  We also use a better frequency resolution for both phases than that quoted in \citet{Ihle2019} (K. Cleary, private communication).

\begin{table}
\centering
\caption{Parameters of COMAP surveys, slightly modified from those quoted in \citet{Ihle2019}. Observing times and survey areas for Phase II are quoted as the value per field times the number of target fields.}
\begin{tabular}{lcc}
\hline
Parameter & COMAP I & COMAP II \\
\hline
System Temperature, $T_{\rm{sys}}$ (K) & 40 & 40 \\
Number of Detectors $N_{\rm{det}}$ & 19 & 95 \\
Beam FWHM, $\theta_{\rm{FWHM}}$ (arcmin) & 4 & 4 \\
Frequency Range, $\nu_{\rm{obs}}$ (GHz) & 26-34 & 26-34 \\
Channel Width, $\delta\nu$ (MHz) & 2.0 & 2.0 \\
Observing Time, $t_{\rm{obs}}$ (hr) & 4500 & 2250$\times$4 \\
Survey Area, $\Omega_{\rm{surv}}$ (deg$^2$) & 1.7 & 2.25$\times$4\\
AGN Source & COSMOS & DESI \\
\hline
\end{tabular}
\label{COMAP_par}
\end{table}

For the noise power spectrum for \ion{C}{II} surveys, we write following \citet{Chung2018b}:
\be
P_N=\frac{\sigma_{\rm{pix}}^2}{t_{\rm{pix}}^{\rm{CII}}}V_{\rm{vox}}^{\rm{CII}}.
\ee
The quantities $t_{\rm{pix}}^{\rm{CII}}$ and $V_{\rm{vox}}^{\rm{CII}}$ are defined by convention slightly differently than their equivalents in the COMAP band, as the solid angle spanned by a single pixel is set to be $\Omega_{\rm{pix}}^{\rm{CII}}=2\pi\sigma_{\rm{beam}}^2$.  For CCATp, CONCERTO, and TIME we use the $\sigma_{\rm{pix}}$ values directly from Table 1 of \citet{Chung2018b}.  For Stage II we take the Noise Equivalent Flux Density from Table 8 of \citet{Silva2015} and convert it to $\sigma_{\rm{pix}}$ by dividing by $\Omega_{\rm{beam}}$.  We can combine these with $V_{\rm{vox}}^{\rm{CII}}$ computed using the modified $\Omega_{\rm{pix}}$ and the frequency bandwidth, and
\be
t_{\rm{pix}}^{\rm{CII}}=\frac{N_{\rm{det}}t_{\rm{obs}}}{\Omega_{\rm{surv}}/\Omega_{\rm{pix}}^{\rm{CII}}}
\ee
to get the noise power.

Table \ref{tab:CII_par} shows the parameters we assume for the \ion{C}{II}-type surveys. Though each has a somewhat different true frequency coverage, we consider a 200-300 GHz frequency band with each instrument for ease of comparison between them.  Stage II parameters are based on \citet{Silva2015}, the others come from \citet{Chung2018b}. As with COMAP , we assume that CCATp can only observe AGN in the 1.7 deg$^2$ COSMOS field, so we reduce its survey area and observing time accordingly.  TIME and CONCERTO plan to observe smaller fields and thus require no modification.  Stage II is a futuristic experiment, so we assume it can cross-correlate with DESI, and thus will see AGN throughout its 100 deg$^2$ footprint.

\begin{table}
\centering
\caption{Parameters of \ion{C}{II} surveys. Stage II parameters are based on \citet{Silva2015}, the others come from \citet{Chung2018b}.}
\setlength\tabcolsep{1.5pt}
\begin{tabular}{lcccc}
\hline
Parameter & CCATp & CONCERTO & TIME & StageII \\
\hline
$\sigma_{\rm{pix}}$ (MJy/sr s$^{1/2}$) & 0.86 & 11.0 & 11.0 & 0.21 \\
$N_{\rm{det}}$ & 20\footnotemark & 3000 & 32 & 16000 \\
$\theta_{\rm{FWHM}}$ (arcsec) & 46.0 & 22.5 & 22.5 & 30.1 \\
$\nu_{\rm{obs}}$ (GHz) & 200-300 & 200-300 & 200-300 & 200-300 \\
$\delta\nu$ (GHz) & 2.5 & 1.5 & 1.9 & 0.4 \\
$t_{\rm{obs}}$ (hr) & 3400 & 1200 & 1000 & 2000 \\
$\Omega_{\rm{surv}}$ (deg$^2$) & 1.7 & 1.4 & 0.01 & 100 \\
AGN Source & COSMOS & COSMOS & COSMOS & DESI \\
\hline
\end{tabular}

\label{tab:CII_par}
\end{table}
\footnotetext{The physical CCATp instrument has $N_{\rm{det}}=1004$. However, the Fabry-Perot inteferometer used for CCATp only observes a fraction of the total spectral range at any given time.  The sensitivity is thus roughly equivalent to that of an effective $N_{\rm{det}}\sim20$ \citep{Chung2018b}}

\subsection{CO Emission Models}
The suite of experiments we consider here are sensitive to several different CO rotational transitions at a variety of redshifts.  Here we summarize the literature models we use to forecast the non-AGN CO emission.  The COMAP surveys primarily target the CO(1-0) transition, while the \ion{C}{II} experiments have the CO(3-2) and higher transitions as ``foregrounds" to their reionization-era \ion{C}{II} signal.  The COMAP surveys are also in principle sensitive to CO(2-1) from the EoR, but we will not consider that line here as large enough spectroscopic quasar samples do not exist at those redshifts.  Table \ref{tab:lines} shows the transitions we consider and the redshift ranges over which we can access them, along with the number density of quasars accessible in the XMM-COSMOS and DESI surveys.  Note that though DESI has a lower density of quasars, it spans a much larger area of the sky, so the total number of quasars available for correlation is much larger than in COSMOS.

\begin{table}
\centering
\caption{Target CO transitions, redshift ranges, and available AGN densities.}
\begin{tabular}{ccccc}
\hline
\multirow{2}{*}{Transition} & \multirow{2}{*}{$z_{\rm{min}}$} & \multirow{2}{*}{$z_{\rm{max}}$} & \multicolumn{2}{c}{AGN/deg$^2$} \\
 & & & XMM-COSMOS & DESI \\
 \hline
(1-0) & 2.4 & 3.4 & 31 & 27 \\
(3-2) & 0.15 & 0.73 & 99 & 4 \\
(4-3) & 0.54 & 1.3 & 200 & 44 \\
(5-4) & 0.92 & 1.9 & 175 & 78 \\
\hline
\end{tabular}
\label{tab:lines}
\end{table}

In order to maintain consistency with past predictions in the literature, we model the non-AGN luminosities of the CO(1-0) line and the higher frequency lines somewhat differently.  For CO(1-0), we use the model from \citet{Li2016}, which assumes a power law relation between the luminosity of a halo and its star formation rate (SFR), and uses the SFR($M,z$) predicted in \citet{Behroozi2013} to get a mass-luminosity relation.  This model also assumes some scatter between the mean $L\subCO(M)$ function and the luminosity of any given halo.  For CO(3-2) and above, we use the model from \citet{Silva2015}, which provides in their Table 4 a direct parameterization of $L\subCO(M)$ for each line.  We neglect the evolution in CO luminosity with redshift, and simply compute power spectra at the center of each frequency band. For all of these models, we use the Tinker mass function and bias \citep{Tinker2008,Tinker2010}, and we compute matter power spectra using CAMB \citep{Lewis2000}.

\begin{figure*}
\centering
\includegraphics[width=\textwidth]{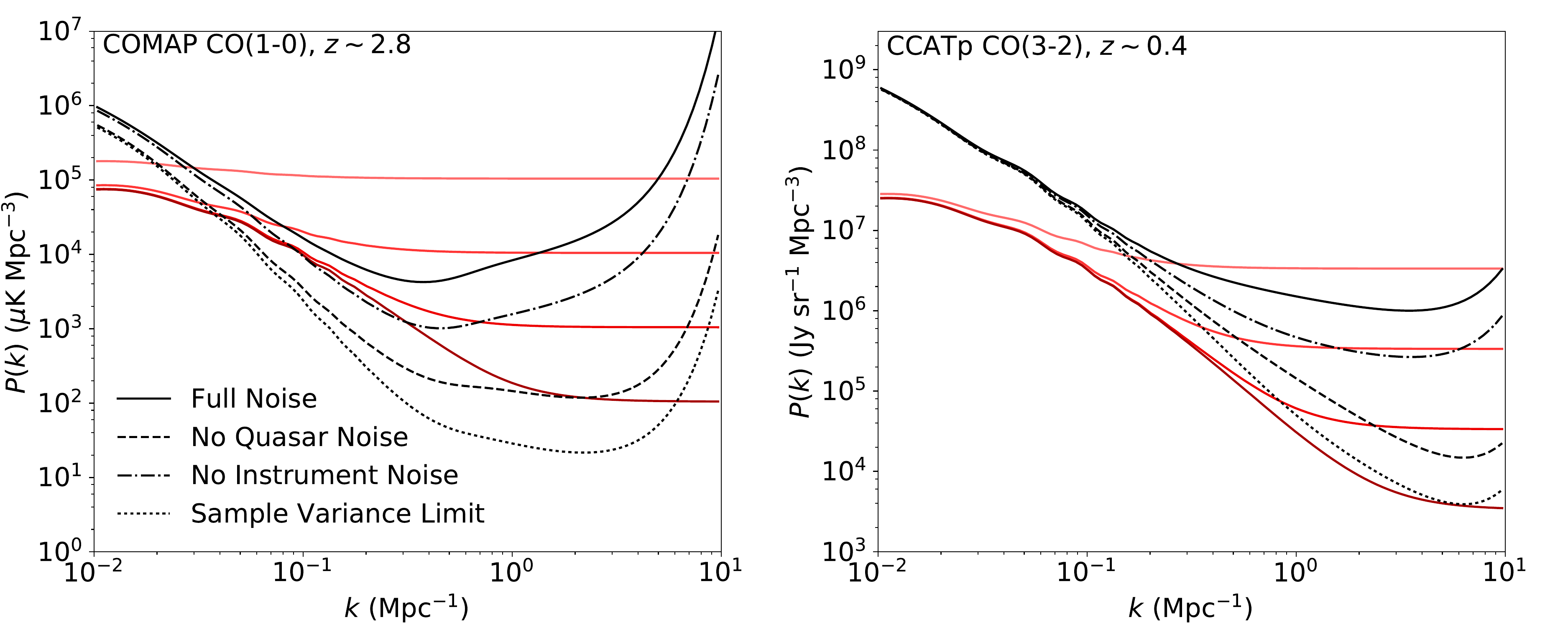}
\caption{CO-AGN cross-spectra for different surveys and different values of $\LQ$.  A model for a COMAP-like survey targeting CO(1-0) at $z\sim3$ is shown in the top panel, one for a CCATp-like survey targeting CO(3-2) at $z\sim0.4$ is shown in the bottom panel.  Red curves show (from dark to light, bottom to top) cross-power spectra for $\log(L_Q/L_{\odot})=5$, 6, 7, 8.  The solid black curves in each panel show the full $\sigma_\times(k)$ computed from Eq. (\ref{sigmaX}), dot-dashed curves show the error without instrumental noise ($P_N=0$), dashed show the error without quasar shot noise ($1/\overline{n}_Q=0$), and dotted curves show the sample-variance limit, where both instrument and quasar shot noise are neglected.  Noise curves are plotted assuming 100 logarithmically spaced $k$ bins between $10^{-2}$ and $10^{1}$ Mpc$^{-1}$.}
\label{fig:Pk}
\end{figure*}

Figure \ref{fig:Pk} shows cross-spectra between CO and AGN for different values of $\LQ$.  The left panel shows spectra for CO(1-0) observed with COMAP, the right CO(3-2) observations with CCATp.  When forecasting the detectability of $\LQ$ with these experiments, we will take $\LQ=0$.  Forecasted errors on $\LQ$ will in this case give the upper limits a given experiment could place on $\LQ$ without a detection.  This is the most conservative assumption we can make given our lack of knowledge of the true $\LQ$ and the fact that we are not modeling the nonlinear clustering power.

\subsection{Scaling with Survey Parameters}
To see how sensitivity to $\LQ$ scales with instrument parameters, we now make some approximations to examine the output of Eq. (\ref{fullFisher}) more closely.  Consider a case where we use the cross-spectrum alone and neglect uncertainties on $\tco b\subCO$ and $b_Q$.  This is equivalent to assuming that the auto-spectra provide sufficient constraints on these quantities enough to avoid cross-spectrum degeneracies.  This assumption turns out to be valid at the ~few percent level for the surveys we consider here, though care must be taken when applying it in general.

Under this assumption, the error on the cross-power can be written as
\be
\sigma_\times(k)=\frac{1}{\sqrt{N_m(k)}}\sqrt{P_\times^2(k)+\left(P_{\rm{CO}}(k)+\frac{P_N}{W^2(k)}\right)P_Q(k)},
\label{sigmaX}
\ee
\citep[][W17]{Villaescusa-Navarro2015}.  Eq. (\ref{sigmaX}) contains contributions from the sample variance of the cross-spectrum and both auto-spectra, along with the shot noise in the CO and AGN maps (implicit in our definitions of $P\subCO$ and $P_Q$), and the instrumental noise in the intensity map.  The black lines in Figure \ref{fig:Pk} show the different contributions to this error, with the full $\sigma_\times$ shown as a solid curve, and errors neglecting instrumental noise and/or quasar shot noise shown as broken curves.

It can clearly be seen from Figure \ref{fig:Pk} that the dominant source of error in both of these measurements is the shot noise from the limited number of AGN.  We can simplify Eq. (\ref{sigmaX}) by taking only terms proportional to $1/\nq$, yielding
\be
\sigma_\times(k)\approx\sqrt{\frac{P\subCO(k) W^2(k)+P_N}{\nq W^2(k)N_m(k)}}
\label{sigmaX_approx}
\ee
As we are neglecting errors on the clustering terms, we only need the Fisher matrix element for $\LQ$ to compute our upper limit:
\be
F_{\LQ\LQ}=\sum_k \frac{1}{\sigma_x^2(k)}\left(\frac{\partial P_\times}{\partial \LQ}\right)^2=\frac{1}{\sigma_{\LQ}^2}.
\label{FLL}
\ee
Combining Eqs. (\ref{sigmaX_approx}) and (\ref{FLL}) with our definitions of the power spectra yields
\be
\sigma_{\LQ}=\frac{1}{\sqrt{f_L\nq}},
\label{sLQ}
\ee
where we have defined
\be
f_L\equiv\clt^2\sum_k\frac{W^2(k)N_m(k)}{P\subCO(k)W^2(k)+P_N}
\label{fL}
\ee

This $f_L$ quantity we have defined encapsulates all of the relevant sensitivity information in the CO survey into a single quantity, including sample variance, shot noise, instrument noise, and resolution effects.  The information in the AGN map is in turn contained entirely in the value of $\nq$.  Eq. (\ref{sLQ}) now provides a simple way to compare the power of different surveys.  Despite all of our simplifying assumptions, it remains accurate at the $\sim$few percent level in comparison to forecasts using the full Fisher matrix from Eq. (\ref{fullFisher}).

\section{Results}
\label{sec:results}

Table \ref{tab:LQ_limits} shows our forecasted 95\% upper limits on $\LQ$ for the surveys we describe above.  In addition to physical $L_{\odot}$ units, we also quote our forecasts in observer-based (K km/s pc$^2$) units.  To make the observer-unit values more readily comparable between different lines, we follow e.g. \citet{Hill2018} (hereafter referred to as H18) and use the CO line ratios measured for QSO's in \citet{Carilli2013} to convert our upper limits to an effective CO(1-0) luminosity $L'_{1-0}$.  This effective luminosity is given by
\be
\frac{L'_{(1-0)}}{\rm{K\ km/s\ pc^2}}\equiv 2.04\times10^5 R_J\left(\frac{\nu_J}{\nu_{(1-0)}}\right)^{-3}\frac{\LQ}{L_{\odot}},
\ee
\citep{Li2016}, where $\nu_J$ is the frequency of the target transition and $R_J$ is the $L`_{(1-0)}/L`_J$ ratio.  We can modify our $f_L$ parameter from Eq. (\ref{fL}) to work in observer units by writing 
\be
f_{L'}\equiv \left(4.9\times10^{-5}\frac{L_{\odot}}{\rm{K}\ \rm{km}/\rm{s}\ \rm{pc}^2}\right)^2 R_J^2\left(\frac{\nu_J}{\nu_{\rm{(1-0)}}}\right)^6f_L
\ee

\begin{table*}
\centering
\caption{95\% Upper Limits on CO luminosity of AGN hosts mapped by different surveys. Values are given as physical luminosities in $L_{\odot}$ and as effective CO(1-0) luminosities in observer units.  Compare to $\overline{L}'_{(1-0)}=4.1\times10^{10}$ K km/s pc$^2$ from directly imaged AGN in H18.}
\begin{tabular}{cccccccccc}
\hline
\multirow{2}{*}{IM Survey} & \multirow{2}{*}{AGN Survey} & \multicolumn{4}{|c}{$\overline{L}_{\rm{Q}}$ (10$^7$ $L_{\odot}$)} & \multicolumn{4}{|c|}{$\overline{L}'_{(1-0)}$ (10$^{10}$ K km/s pc$^2$)} \\
&  & 1-0 & 3-2 & 4-3 & 5-4 & 1-0 & 3-2 & 4-3 & 5-4 \\
\hline
COMAP I & zCOSMOS & 0.18 & -- & -- & -- & 3.6 & -- & -- & -- \\
CCATp & zCOSMOS & -- & 1.1 & 3.7 & 9.7 & -- & 0.94 & 1.4 & 2.3 \\
CONCERTO & zCOSMOS & -- & 0.69 & 2.2 & 5.7 & -- & 0.60 & 0.80 & 1.3 \\
TIME & zCOSMOS & -- & 10.0 & 31.4 & 82.0 & -- & 8.7 & 11.5 & 19.4 \\
\hline
COMAP II & DESI & 0.063 & -- & -- & -- & 1.2 & -- & -- & -- \\
Stage II & DESI & -- & 0.003 & 0.005 & 0.009 & -- & 0.003 & 0.002 & 0.002 \\
\hline
\end{tabular}
\label{tab:LQ_limits}
\end{table*}

To judge the usefulness of our forecasted measurements, we can compare the upper limits in Table \ref{tab:LQ_limits} to values obtained through direct observations of CO in quasars.  For example, we can compare our results to the 13 QSOs observed in CO(3-2) by H18. As currently observations of individual quasars in molecular gas can be very expensive, many objects chosen for observation have been previously detected in the sub-mm or are in some way extreme. This holds for the objects observed in H18 which are considered to be rare, hyper-luminous and UV-bright QSOs at $2.5 < z < 2.9$ which, additionally, were all previously detected in the sub-mm with SCUBA-2. Thus, we might expect the actual value of $\LQ$ to be substantially lower than this value.

H18 find an average CO(3-2) luminosity of $\overline{L}'_{(1-0)}=4.1\times10^{10}$ K km/s pc$^2$. As shown in H18 figure 5, using the average value of $\overline{L}'_{(1-0)}$ from H18 is a good indicator of the typical value currently observed in high-z ($\gtrsim 1.4$) quasar samples. The observed values for low-z AGN ($\lesssim 0.4$) could be an order of magnitude lower \citep{Scoville2003}. Nevertheless, we see little redshift evolution in our cross-correlation sensitivity and most of our results probe samples at $z>0.4$ so we find H18 to be an appropriate comparison. Thus we can use this value of $\overline{L}'_{(1-0)}$ as a benchmark for comparing the quality of the surveys we consider here. In other words, if all quasars had CO luminosities similar to the H18 sample, we would expect to detect cross-shot power at reasonable significance in CCATp, CONCERTO, and COMAP II, along with an extremely precise measurement in \ion{C}{II} StageII.  The COMAP Phase I survey in our forecast has enough sensitivity to marginally detect the H18 quasars, and the TIME survey appears to be not well-optimized for this type of observation.

\begin{figure*}
\includegraphics[width=0.8\textwidth]{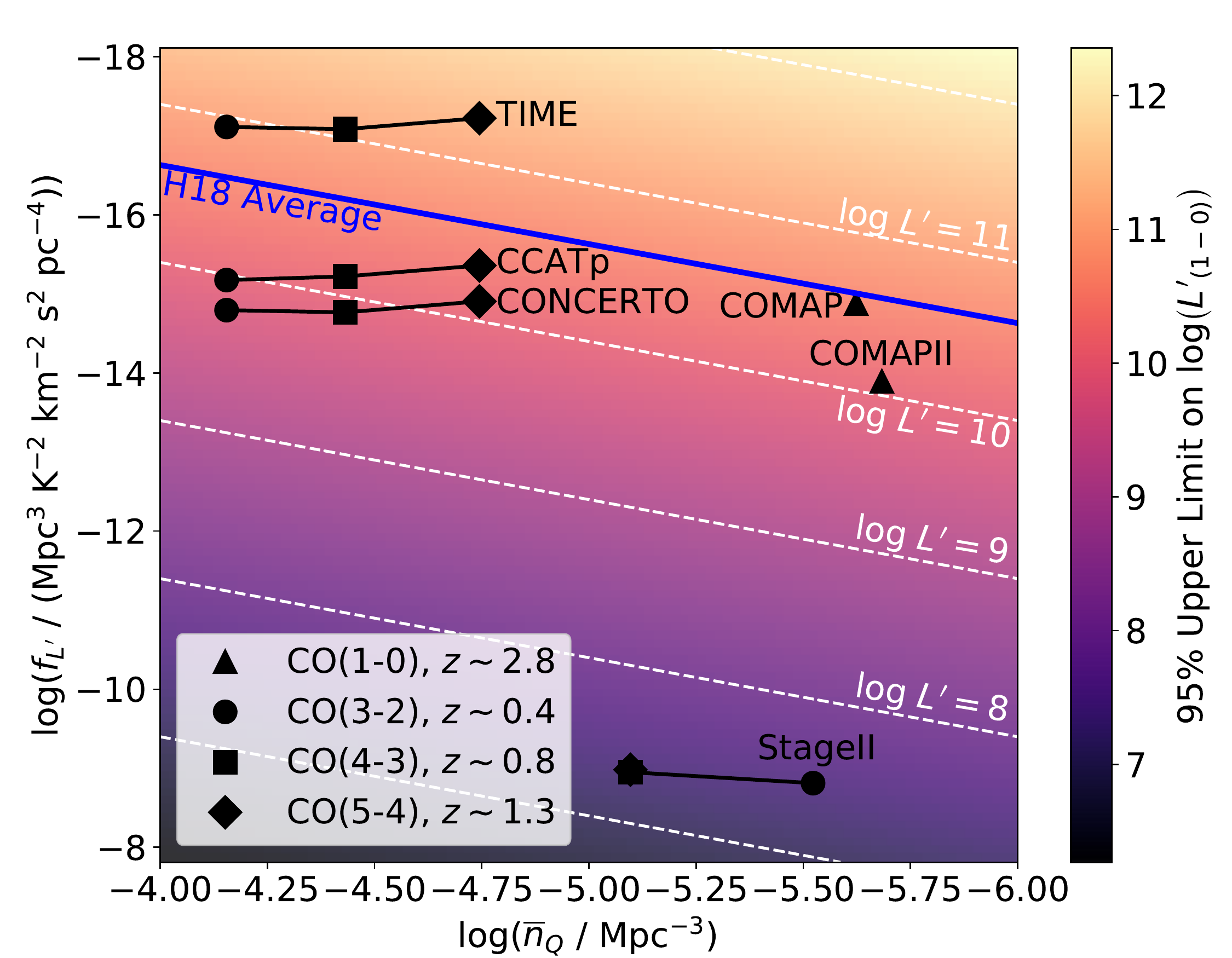}
\caption{Forecasted 95\% upper limit on $\overline{L}'_{(1-0)}$ as a function of quasar number density $\nq$ and CO survey quality $f_{L'}$.  Contours showing different values of $\overline{L}'_{(1-0)}$ are shown as white dashed lines, while the blue solid line shows the average value from the H18 quasars.  Values for different experiments are shown as labeled black markers with lines connecting limits for different transitions observed by the same experiment.  Triangles show forecasts from CO(1-0) observations, circles from CO(3-2), squares from CO(4-3), and diamonds from CO(5-4).}
\label{fig:fLnQ}
\end{figure*}

To illustrate more clearly the dependence of our forecast on survey design, Figure \ref{fig:fLnQ} shows the upper limit a survey could obtain on $\overline{L}'_{(1-0)}$ as a function of $\nq$ and the $f_L$ parameter defined in Eq. (\ref{fL}).  Though this only approximates the values used in Table \ref{tab:LQ_limits}, as we stated previously the approximation should be valid at the $\sim$few percent level.  Plotting the forecasts in this way allows us to completely separate out the dependence on the AGN survey and the CO survey.

Of the current-generation surveys, we find that CCATp and CONCERTO can make the most sensitive measurements of $\LQ$.  CONCERTO does slightly better than CCATp, largely because it has higher frequency resolution, making it more sensitive to the small spatial scales we care about here.  The TIME survey does not do as well as the other \ion{C}{II} experiments as it covers a much smaller volume, limiting it to a small number of Fourier modes and relatively few quasars.  COMAP Phase I, because it observes at higher redshift than the \ion{C}{II} experiments, also has fewer AGN to work with and less sensitivity to small scales.  The latter is compounded by the fact that COMAP targets the lowest-frequency CO line.  However, the added sensitivity of COMAP Phase II/DESI is enough to significantly improve the Phase I constraints.  The \ion{C}{II} Stage II experiment is by far the most ambitious experiment we consider here.  Because its high sensitivity allows it to map an order of magnitude more sky than the other experiments, it has a much lower $f_L$ and can measure $\LQ$ at extremely high precision.

The forecasts for the three current-generation \ion{C}{II} experiments evolve similarly with the targeted CO transition.  In the COSMOS field, the total number of AGN per sky area is highest in the CO(4-3) band, while the total comoving volume of the surveys decreases monotonically to higher $J$'s.  This causes the comoving quasar density to decrease for higher-order lines, and worsens the forecasted limits.  The $f_L$ parameter for these surveys also gets slightly worse with increasing $J$, though at slightly different rates for the three surveys as they have very different parameters.  For the Stage II survey, the constraints improve significantly between CO(4-3) and CO(5-4).  This is because, unlike in COSMOS, the number of AGN per square degree actually increases from CO(4-3) to CO(5-4).  The comoving number density  remains roughly constant between these two lines, leading to the observed behavior in Figure \ref{fig:fLnQ}.

\section{Science Applications}
\label{sec:science}

The CO cross-power spectrum has strong implications for the study of molecular gas in quasars and we envision several possible science applications.  Of course, our ability to extract scientifically-interesting information from this sample is influenced by our ability to convert from $\overline{L}_{\rm{Q}}$ to a quantity of scientific interest, specifically the gas mass, or $M_{\rm{gas}}$. There is disagreement over estimates for this conversion factor, $\alpha_{\rm{CO}}$ for AGN hosts \citep{Bolatto2013} with some papers adopting the Milky Way value, $\alpha_{CO(1-0)} \approx 4$M$_{\bigodot}$(K km s$^{-1}$ pc$^2$)$^{-1}$ \citep{Feruglio2010,Herrera2019} while others use the value typical of ULIRGs $\alpha_{CO(1-0)} \approx 0.8-1.0$M$_{\bigodot}$(K km s$^{-1}$ pc$^2$)$^{-1}$ \citep{Cicone2014,Fiore2017,Veilleux2017,Hill2018,Fleutsch2019}.  Measurements to confirm this value for a relevant sample of AGN, will be necessary for proper interpretation of our results.

If the mean CO luminosity of DESI quasars is around the H18 level, then under our assumptions a survey like \ion{C}{II} StageII will provide sub-percent level measurements of the mean CO luminosity in AGN hosts.  At that point, our understanding of the details of AGN feedback would likely be limited by the precision of our models, as detailed computation of the physics which determines CO emission is a highly complicated task (see, e.g. \citealt{Popping2019}).  However, this level of precision (and the excellent redshift resolution offered by intensity mapping) means that one can break the full frequency bandwidth of the map into several shells.  This will sacrifice sensitivity in each band, but could pay substantial dividends by allowing one to map the detailed redshift evolution of AGN feedback processes and search for changes in the average gas mass of AGN hosts as a function of redshift.

Several of the proposed \ion{C}{II} experiments will, in fact, give access to several different CO transitions, and thus allow us to explore the CO spectral line energy distribution (SLED) for quasars (assuming a proper accounting for the redshift evolution of the average CO luminosity of quasars as mentioned above).  There is emerging evidence to suggest that quasars may have more highly excited molecular gas than submillimter galaxies (SMGs) or main sequence galaxies.  \citet{Riechers2011} find evidence for a single, highly excited gas component in a sample of five redshift $z \sim 2.5$ lensed quasars. Evidence suggests that all of the molecular gas is associated with star formation.  In addition, it is possible that either wind- or jet-mediated quasar feedback would introduce a second more excited component of molecular gas.  For example, \citet{Dasyra2016} found that the ratio of CO(4-3)/CO(2-1) in IC 5063 was $\approx 1$  for the disk but closer to 5 in regions near the AGN-driven jet (necessitating that the gas be optically thin). Our technique would allow us to probe the average quasar SLED which will facilitate the comparison of future observations made at different CO transitions and reveal details of the excitation state of the molecular gas in AGN hosts.

While this paper explores the possible result of cross correlating CO intensity mapping experiments with AGN samples, this technique is applicable for any number of galaxy samples. Leaving aside for now the science questions to be answered by different galaxy samples, it will also be possible to compare the results of the CO cross-power spectrum for quasar and other galaxy samples. This is where we see the most power to shed insight on the question of quasar feedback. Comparing the CO cross-power spectrum of AGN hosts to a mass-matched sample of star-forming and quenched galaxies will allow us to search for a difference in the mean CO luminosity. As the samples are independently-selected on small scales where the shot noise term dominates the power spectrum, these results will be dominated by the measurement error on the smaller quantity, expected to be the quasar sample due to their relative lack of abundance. Thus, the measurement errors on our ability to calculate the difference in $\overline{L}_{\rm{CO}}$ between a sample of quasars and a complimentary sample of galaxies will be similar to those expressed in Table \ref{tab:LQ_limits}. 

If AGN are in fact responsible for clearing out the molecular gas from their hosts, the depletion timescale is expected to be on the order of a few hundred Myr \citep{Cicone2014, Fiore2017}. Thus, we would expect the average gas mass in the AGN hosts to be comparable to that of a mass-matched sample of passively evolving galaxies and significantly gas-deficient compared to a mass-matched sample of star-forming galaxies (alternatively, one could compare to a population of galaxies selected to have the same large-scale linear bias as the AGN sample).  Such results have not been confirmed with current AGN samples \citep[see, for example][]{Shangguan2018} though there does appear to be a dearth of molecular gas in quasars as compared to star-forming galaxies as a function of stellar mass if accounting for the redshift evolution of such trends for star-forming galaxies \citep{Genzel2015,Fiore2017}. Alternatively, some observations suggest that swept-up gas from AGN outflows may produce additional molecular material \citep{Richings2018} in which case quasar samples may have higher gas masses than any comparison sample. Our calculations show that upcoming intensity mapping surveys will allow us to distinguish between these two scenarios.

Finally, the search for feedback from AGN also rests on the presence of molecular outflows as measured by an excess of blueshifted emission in the CO transitions of AGN hosts \citep{Cicone2014, Veilleux2017}.  In an instrument with sufficient frequency resolution, outflows would create an extra source of small-scale anisotropy in our power spectra.  This signature comes from the fact that the CO emission in quasars would be extended in frequency space while remaining compact in the plane of the sky.  Searches for quasar outflows tend to look for gas with outflow speeds $\gtrsim 500$ km/s \citep{Cicone2014}. The frequency resolution of COMAP or a Stage II \ion{C}{II} experiment would be sufficient to resolve gas at such speeds and thus we would expect to be able to detect such a signature in CO-AGN cross correlations. A determination and detectability forecast of the average quasar outflow speed is beyond the scope of this paper but will be discussed in future work.  As with the other probes we can compare cross-correlations of CO with AGN with correlations between CO and star forming galaxies to search for specifically quasar-driven outflows. 

\section{Discussion}
\label{sec:discussion}

We described here the potential of correlating CO and AGN surveys to study AGN feedback.  This is far from the only useful science which can be gleaned from cross-correlations using the surveys we discuss here.  We discussed correlating star forming and quiescent galaxies in COSMOS and DESI with our CO intensity maps as a means of comparing molecular gas properties between galaxies which do and do not host AGN.  However, these additional correlations are interesting in their own right, and could be used to study how molecular gas mass scales with a variety of different galaxy properties in a similar manner to the work done in W17.  One could also cross-correlate with other surveys.  For example, the correlation between COMAP and the Hobby-Eberly Telescope Dark Energy Experiment (HETDEX, \citealt{Hill2008}) discussed in \citet{Chung2019} can be used to study the CO emission from Lyman~$\alpha$ emitters.  Beyond that, the cross-spectra we consider here only use a small fraction of the vast DESI survey area.  All-sky intensity maps like those produced by the upcoming Spectro-Photometer for the History of the Universe, Epoch of Reionization, and Ices Explorer satellite (SPHEREx, \citealt{Dore2014}) will be able to correlate with DESI to produce results integrating over orders of magnitude more galaxies than we consider here, opening an unprecedented window into the high-redshift universe.

One way this cross-correlation method can potentially be improved would be if one could make effective use of photometric galaxy surveys.  Typical number densities of AGN with photo-$z$ measurements are around an order of magnitude higher than spectroscopic quasars.  The exception here may be COSMOS due to the extensive spectroscopic follow-up of AGN candidates. Only about 50\% of photometrically-classified AGN have not been confirmed with follow-up spectroscopy in this field. On the other hand, the upcoming Large Synoptic Survey Telescope \citep[LSST;][]{Ivezik2008} will identify approximately $10^7$ AGN \citep{LSST2009}, with a significant fraction at $z> 2.4$ but will have only photometric redshifts for the vast majority of these objects.  As quasar shot noise is the dominant source of error in our forecasts, using photo-$z$ AGN would na\"ively seem to dramatically improve the power of our forecasts, as can be seen in Figure \ref{fig:fLnQ}.  However, \citet{Chung2019} demonstrated that decreasing the redshift accuracy of a galaxy survey to photometric levels significantly reduces the strength of the correlation, implying that some of this gain in sensitivity would be lost.  Beyond raw statistical significance, this also would add a potentially difficult systematic to the measurement, as one would need to know the true redshift distribution of the photo-$z$ AGNs to reconstruct what the true cross-shot noise would be absent this signal loss.  We leave for future work the question of whether the increased quasar number counts are worth dealing with these challenges.

We consider COSMOS here as a cross-correlation field for COMAP Phase I, but \citet{Chung2019} found that cross-correlating with HETDEX would be a better option for overall cross-correlation signal to noise.  Specifically, it was found that the best signal-to-noise would come from targeting the Spitzer/HETDEX Exploratory Large-Area field (SHELA), which covers $\approx$24 degrees of the Sloan Digital Survey ``Stripe 82" \citep{Annis2014}. HETDEX is designed to pick out Lyman~$\alpha$ emitters between $1.9 < z < 3.5$ using wide-field integral field unit (IFU) spectrographs and so will undoubtedly contain a large number of AGN with spectroscopic redshifts for which a similar cross-correlation can be conducted. Estimates for the exact number density currently vary though assuming a similar detection rate from their test field \citep{Hill2008} of one AGN per 40 arcmin with a filling factor of 1/3 implies a number density of $\sim$five AGN/deg$^2$ in the SHELA field where the filling factor will be of order unity \citep{Papovich2016}.

We have neglected in this paper any discussion of continuum foregrounds in our CO maps.  This is perhaps the most significant challenge facing 21 cm intensity mapping surveys, as foreground contamination at low frequencies is several orders of magnitude brighter than the cosmological signal (see for example \citealt{Liu2011}).  Continuum foregrounds are less of a challenge in maps of other lines, but they can be dealt with, as they are much less dominant over the signal and largely remain contained to the lowest line-of-sight Fourier modes \citep{Keating2015}.  Even if residuals remain after foreground cleaning, cross-spectra like the ones we propose here are very robust against foreground contamination, as the cosmological signal will be spatially correlated with the AGN locations and the foregrounds generally will not be, as demonstrated in \citet{Switzer2019} and \citet{Cunnington2019}.

\section{Conclusion}
\label{sec:conclusion}
In this paper, we have demonstrated that a cross-correlation between a CO intensity map and a spectroscopic quasar survey can provide valuable new insights into the process of AGN feedback.  Current or near-future intensity mapping experiments such as COMAPII, CCATp, and CONCERTO will have sufficient sensitivity to study the molecular gas content of AGN hosts with far less bias than current directed observations. This will provide crucial insight in to the effect of ``quasar mode" feedback on the molecular reservoirs of a uniformly-selected sample of AGN hosts, where the selection process can be done independently of knowledge of the molecular content or star formation rate of the host but solely as a function of AGN luminosity. Such a measurement will allow us to determine, with robust statistics, the effect of this feedback on the gas mass of the host galaxy and shed insight on to contradictory evidence for negative and positive feedback in AGN hosts. Comparing quasar samples with similar samples of star forming and quiescent galaxies will allow us to determine whether the molecular reservoirs of AGN hosts more closely resemble those of actively star-forming galaxies or quiescent galaxies with exhausted molecular reservoirs.

Future experiments like \ion{C}{II} StageII and DESI will dramatically improve on these near-term measurements, offering deep maps over large enough volumes to study the molecular gas content of AGN hosts in exquisite  detail. In particular, larger samples of quasars paired with large area CO intensity mapping surveys will enable us to chart the evolution in the gas mass of quasars as a function of redshift or other relevant parameters, and to search for evidence of molecular outflows with speeds $> 500$ km/s. Such measurements will allow us to directly view the impact of the AGN population on their host galaxies and search for evidence of the depletion of molecular material in their hosts due to molecular outflows. Molecular outflows specifically can be a ``smoking gun" signal of AGN feedback impacting the host galaxy, making future intensity mapping experiments with higher frequency resolution highly desirable for cross-correlation. 

We have highlighted here one of many exciting applications for cross-correlations between galaxy surveys and intensity maps.  As increasing amounts of data become available, this and other similar analyses will open fascinating new windows into the physics of distant galaxies and expand the research possibilities available because of intensity mapping surveys.

\section*{Acknowledgements}
We thank Alberto Bolatto, Kieran Cleary, Dongwoo Chung, Neal Dalal, Hamsa Padmanabhan, Tina Peters, Anthony Pullen, and Eric Switzer for useful discussions.  R.M.A. would like to acknowledge funding from the National Science and Engineering Research Council of Canada.

\bibliography{references}

\appendix
\section{Power Spectra Derivations}
\label{app:Pk_derivation}
Here we will derive the relevant auto- and cross-spectra between our CO and AGN maps, loosely following the derivation from W17.  As these derivations are rarely found in the literature, we present them here in full detail for the convenience of the reader.

\subsection{AGN Auto Power}
First consider a map containing the locations of $N_Q$ quasars in volume $V$ (this derivation is true for any discrete galaxy survey, but we focus on quasars here as that is what we use above).  To derive the power spectrum of such a map, we first split the volume into $N_c$ infinitesmial cells with small volume $\delta V$ such that cell $i$ contains $N_i=0$ or $1$ quasars.  We can then take advantage of the useful property that $N_i^2=N_i$.  The power spectrum $P_Q(k)$ of the quasar density contrast $\dq$ is defined by the estimator
\be
P_Q(k) = V\left<\left|\dkq(\K)\right|^2\right>,
\label{eq:Pgal}
\ee
where $k$ is the magnitude of Fourier wavenumber $\K$.  The tilde indicates a Fourier transform
\be
\dkq(\K)=\frac{1}{V}\int\dq(\X)e^{\im\K\cdot\X}d^3\X=\frac{1}{N_Q}\sum_{i=1}^{N_c}N_ie^{\im\K\cdot\X_i},
\ee
where in the second equality we have replaced the integral with a sum over our infinitesimal cells (note that the second equality is only valid for $\K\neq0$).

We can then write
\be
\left<\left|\dkq(\K)\right|^2\right>=\frac{1}{N_Q^2}\sum_{i,j=1}^{N_c}\left<N_iN_j\right>e^{\im\K\cdot(\X_i-\X_j)}
\label{eq:gal1}
\ee
This sum can be divided into two parts, one where $i\neq j$, i.e. the contribution from pairs of sources, and one where $i=j$, the contribution from each source correlated with itself.  In the first case, we have 
\begin{multline}
\cg_{i\neq j}=\frac{1}{N_Q^2}\sum_{i\neq j}\nq^2\delta V^2\left[1+\xi_Q(\X_j-\X_i)\right] \\ \times\exki,
\label{eq:cfq}
\end{multline}
where $\xi_Q$ is the two-point correlation function of the quasar population.  If we assume a simple linear bias $b_Q$, this is related to the correlation function of the underlying dark matter field $\xi_m$ by $\xi_Q=b_Q^2\xi_m$.  We can rewrite Eq. (\ref{eq:cfq}) as
\begin{multline}
\cg_{i\neq j}=\frac{1}{V^2}\int\exkp d^3\X d^3\X' \\ +\frac{1}{V^2}b_Q^2\int\xi_m(\X-\X')\exkp d^3\X d^3\X'.
\end{multline}
The first integral vanishes due to symmetry, and we can simplify the second by recognizing that statistical isotropy and homogeneity requires that $\xi_m$ depend only on $\R\equiv\X-\X'$.  This yields
\be
\cg_{i\neq j}=\frac{b_Q^2}{V^2}\int\xi_m(\R)e^{-\im\K'\cdot\R}d^3\R,
\ee
which is simply the definition of the matter power spectrum $P_m(k)$.  Thus we have
\be
\cg_{i\neq j}=\frac{1}{V}b_Q^2P_m(k).
\label{eq:gal2}
\ee

For the $i=j$ component, we can instead rewrite Eq. (\ref{eq:gal1}) as
\be
\cg_{i=j}=\frac{1}{N_Q^2}\sum_{i=1}^{N_c}\left<N_i^2\right>e^{\im\K\cdot\X}.
\ee
Since $\left<N_i^2\right>=\left<N_i\right>=\nq\delta V$, we have
\be
\cg_{i=j}=\frac{\nq}{N_Q^2}\sum_{i=1}^{N_c}\delta V.
\ee
The sum of the volumes of all the cells by definition must add up to $V$, so we have
\be
\cg_{i=j}=\frac{1}{N_Q}.
\label{eq:gal3}
\ee

Combining Eq's (\ref{eq:gal2}) and (\ref{eq:gal3}) with Eq. (\ref{eq:Pgal}) yields the full form of the quasar power spectrum:
\be
P_Q(k)=b_Q^2P_m(k)+\frac{1}{\nq},
\ee
with clustering and shot-noise components identical to those given in Eq. (\ref{galauto}).

\subsection{CO Auto Power}
Next we consider the power spectrum of our CO intensity map.  We again divide our map into infinitesmial cells, but we now will start by considering the intensity contributed only by halos with CO luminosities between $L$ and $L+dL$.  The intensity in cell $i$ from these halos is
\be
T(\X_i,L)dL=\clt LN_i(L)dL,
\ee
where $N_i(L)=0$ or 1 is the number of emitters with the given luminosity and $\clt$ is the conversion factor defined in Eq. (\ref{CLT}).  The power spectrum of an intensity map is defined similarly to that of a galaxy map, through
\be
P_{\rm{CO}}(k)\equiv V\ca.
\label{eq:CO1}
\ee

Now we need to break our estimator again into contributions from pairs of sources and from single sources, and also down into sources with different luminosities.  Starting with the two-source part, we have
\begin{multline}
\caL_{i\neq j}dL_1dL_2=\clt^2\frac{L_1L_2}{V^2} \\ \times\sum_{i\neq j}\left<N_i(L_1)N_j(L_2)\right>\exki dL_1dL_2.
\end{multline}
Similar to before, the expectation value $\left<N_i(L)\right>=dn/dL \delta_V$, where $dn/dL$ is the luminosity function of the CO line.  We can also define a bias $b(L)$ between the dark matter density and the CO emitter density.  We then can write
\begin{multline}
\caL_{i\neq j}dL_1dL_2=\clt^2\frac{L_1L_2}{V^2}\frac{dn}{dL_1}\frac{dn}{dL_2} \\ \times\sum_{i\neq j}\delta V^2\left[1+b(L_1)b(L_2)\xi_m(\X_i-\X_j)\right] \\ \times\exki dL_1dL_2.
\end{multline}
This can be simplified following the same logic used to derive Eq. (\ref{eq:gal2}), yielding
\begin{multline}
\caL_{i\neq j}dL_1dL_2=\frac{\clt^2}{V}L_1b(L_1)\frac{dn}{dL_1} \\ \times L_2b(L_2)\frac{dn}{dL_2}P_m(k).
\end{multline}
Finally, integrating over all source luminosities gives
\be
\ca_{i\neq j}=\frac{\clt^2}{V}\left[\int Lb(L)\frac{dn}{dL}dL\right]^2P_m(k),
\label{eq:CO2}
\ee
where we have assumed that $L_1$ and $L_2$ are uncorrelated, an assumption which should hold on large scales.

For the single-source term, we have
\begin{multline}
\caL_{i=j}dL_1dL_2=\clt^2\frac{L_1L_2}{V^2}\\ \times\sum_{i=1}^{N_c}\left<N_i(L_1)N_i(L_2)\right>e^{\im\K\cdot\X_i}dL_1dL_2.
\end{multline}
Taking advantage again of the properties of $N_i$, we can write that
\begin{eqnarray}
\left<N_i(L_1)N_i(L_2)\right>&=&\left<N_i(L_1)\right>\delta_D(L_2-L_1) \\ \nonumber
&=&\frac{dn}{dL_1}\delta V\delta_D(L_2-L_1),
\end{eqnarray}
where in this case the Dirac delta function enforces that $L_1=L_2$ since both describe the same source.  We thus have
\begin{multline}
\caL_{i=j}dL_1dL_2=\clt^2\frac{L_1L_2}{V^2} \\ \times \delta_D(L_2-L_1)\frac{dn}{dL_1}\sum_{i=1}^{N_c}\delta Ve^{\im\K\cdot\X}dL_1dL_2,
\end{multline}
which simplifies to
\begin{multline}
\caL_{i=j}dL_1dL_2=\frac{\clt^2}{V}L_1L_2\frac{dn}{dL_1}\\ \times\delta_D(L_2-L_1)dL_1dL_2.
\end{multline}
Integrating over luminosities gives
\be
\ca_{i=j}=\frac{\clt^2}{V}\int L^2\frac{dn}{dL}dL.
\label{eq:CO3}
\ee

Combining Eq.'s (\ref{eq:CO2}) and (\ref{eq:CO3}) with (\ref{eq:CO1}) gives the full form of the intensity mapping power spectrum from Eq. (\ref{COauto}):
\be
P_{\rm{CO}}(k)=\overline{T}_{\rm{CO}}^2b_{\rm{CO}}^2P_m+P^{\rm{shot}}_{\rm{CO}},
\ee
where $\overline{T}_{\rm{CO}}$, $b_{\rm{CO}}$, and $P^{\rm{shot}}_{\rm{CO}}$ have their definitions from Section \ref{sec:method}.

\subsection{CO/AGN Cross Power}
Finally, we want to compute the cross-power spectrum $P_\times(k)$ between our two observables, defined via
\be
P_\times(k)\equiv V\left<\tilde{T}(\K)\tilde{\delta}_Q^*(\K)\right>.
\label{eq:x1}
\ee
For our purposes, we will assume that the biases for both tracers have no scale dependence.  We follow the same procedure as before.  Consider first the two-source component, for CO emitters with luminosities between $L$ and $L+dL$,
\begin{multline}
\cxL_{i\neq j}dL=\clt\frac{L}{VN_Q}\\ \times\sum_{i\neq j}\left<N_i(L)N_j\right>\exki dL.
\end{multline}
The volume has still been split into infinitesimal cells, $N_i(L)$ in this case refers to the number of CO emitters with luminosity between $L$ and $L+dL$ in cell $i$, $N_j$ refers to the number of quasars in cell $j$.  Both $N_i(L)$ and $N_j$ can still only be 0 or 1.  From here, we proceed similarly to the previous two cases:
\begin{multline}
\cxL_{i\neq j}dL = \clt\frac{L}{VN_Q}\frac{dn}{dL}\nq \\ \times\sum_{i\neq j}\delta V^2\left[1+b(L)b_Q\xi_m(\X_j-\X_i)\right]\exki dL
\end{multline}
\be
\cxL_{i\neq j}dL = \frac{\clt}{V}Lb(L)\frac{dn}{dL}b_QP_m(k)dL,
\ee
with the final average over CO luminosities yielding
\be
\cx_{i\neq j}=\frac{\clt}{V}\left[\int Lb(L)\frac{dn}{dL}dL\right]b_QP_m(k).
\ee

The calculation of the contribution from single sources also begins similarly to before:
\begin{multline}
\cxL_{i=j}dL=\clt\frac{L}{VN_Q} \\ \times\sum_{i=1}^{N_c}\left<N_i(L)N_i\right>e^{\im\K\cdot\X_i}dL
\end{multline}
Terms in the sum are now only nonzero if $N_i(L)$ and $N_i$ are both 1, i.e. if the source in a cell is both a quasar and a CO emitter.  Thus we have
\be
\left<N_i(L)N_i\right>=\qlf\delta V,
\ee
where $dn/dL|_Q$ is the CO luminosity function of only the directly-observed quasars.  Then,
\begin{eqnarray}
\cxL_{i=j}dL&=&\frac{\clt L}{VN_Q}\qlf\sum_{i=1}^{N_c}e^{\im\K\cdot\X_i}dL \nonumber \\
&=&\clt\frac{L}{N_Q}\qlf dL.
\end{eqnarray}

The full cross-spectrum is then, after integrating over luminosity,
\be
P_\times(k)=\overline{T}_{\rm{CO}}b_{\rm{CO}}b_QP_m(k)+\clt\overline{L}_Q,
\ee
which is the form given in Eq. (\ref{Px}).
\end{document}